\documentclass[12pt]{article}
\usepackage{layout,amssymb,amsmath,epsfig}

\begin{document}
\title{\bf Energy Conditions in $f(R,T,R_{\mu\nu}T^{\mu\nu})$ Gravity}
\author{M. Sharif \thanks{msharif.math@pu.edu.pk} and M.
Zubair \thanks{mzubairkk@gmail.com}\\
Department of Mathematics, University of the Punjab,\\
Quaid-e-Azam Campus, Lahore-54590, Pakistan.}

\date{}

\maketitle

\begin{abstract}
We discuss the validity of the energy conditions in a newly modified
theory named as $f(R,T,R_{\mu\nu}T^{\mu\nu})$ gravity, where $R$ and
$T$ represent the scalar curvature and trace of the energy-momentum
tensor. The corresponding energy conditions are derived which appear
to be more general and can reduce to the familiar forms of these
conditions in general relativity, $f(R)$ and $f(R,T)$ theories. The
general inequalities are presented in terms of recent values of
Hubble, deceleration, jerk and snap parameters. In particular, we
use two specific models recently developed in literature to study
concrete application of these conditions as well as Dolgov-Kawasaki
instability. Finally, we explore $f(R,T)$ gravity as a specific case
to this modified theory for exponential and power law models.
\end{abstract} {\bf Keywords:} $f(R,T,R_{\mu\nu}T^{\mu\nu})$ gravity; Raychaudhuri equation;
Energy conditions.\\
{\bf PACS:} 04.50.-h; 04.50.Kd; 98.80.Jk; 98.80.Cq.

\section{Introduction}

In current scenario, dark energy (DE) is referred as an active agent which
tends to accelerate the expansion in cosmos. The expanding paradigm of the
universe has been affirmed from various observational measurements \cite{1}.
Modified theories have received much attention to count with the issue of
cosmic acceleration. In these theories, modified gravity models have been
formulated to recognize the origin of DE as modification to the
Einstein-Hilbert action. One of the fascinating class of models is named as
$f(R)$ gravity where the generic nonlinear function $f$ succeeds the Ricci
scalar (for review see \cite{2,3}). In this theory, one can reproduce the
cosmological constant scenario, i.e., the classic $\Lambda$ cold dark matter
($\Lambda$CDM) model by choosing the $f(R)$ Lagrangian as $f=R-2\Lambda$. The
$\Lambda$CDM expanding paradigm has been tested and confirmed from the recent
results \cite{4}. One interesting fact about this theory is that it develops
equivalence with the Brans-Dicke (BD) \cite{5} theory for a specific BD
parameter. The BD theory involves nonminimal coupling between geometry and
scalar field which has further been formulated in $f(R)$ theory \cite{6,7}.

Bertolami et al. \cite{7} put a new twist on $f(R)$ gravity by considering
the Lagrangian as a function of scalar curvature explicitly coupled with
matter Lagrangian density. Bertolami and Paramos \cite{8} developed the
correspondence between this modified theory and scalar-tensor theory in which
nonminimal curvature matter coupling would yield two scalar fields. The
interaction between the matter components and curvature terms results in non
conservation of matter energy-momentum tensor \cite{9} which may describe the
cosmic acceleration \cite{10}. Later, Harko \cite{11} extended this theory by
inserting a general function of matter Lagrangian. Nesseris \cite{12} studied
matter density perturbations to constrain this theory from growth factor as
well as weak lensing observations. Wu \cite{12a} established the laws of
thermodynamics in this modified theory and discussed some forms of curvature
components. Harko and Lobo \cite{13} suggested a more generalized form of
$f(R)$ theory by taking Lagrangian as a generic function of $R$ and matter
Lagrangian $\mathcal{L}_m$.

In modified $f(R)$ theories of the type involving nonminimal coupling with
matter Lagrangian suffer issue related to the choice of matter Lagrangian
density. If $\mathcal{L}_m=p$ is considered then extra force would be
vanished out so that natural conservation of matter exist in such case
\cite{14}. One can still get the effective nonminimal coupling if
$\mathcal{L}_m=-\rho$ \cite{15}. Another way of modifying the Einstein
Lagrangian is to consider the function of trace of the energy-momentum tensor
$T$ \cite{16} such that $\Lambda$CDM model can be considered of the form
$R+2\Lambda(T)$. Harko et al. \cite{17} implemented this idea to generate a
new Lagrangian $f(R,T)$ where the matter geometry coupled system is
introduced with arbitrary function of $R$ and $T$. The corresponding
effective matter geometry coupling favors the non-geodesic motion of test
particles leading to extra force as suggested in other modified theories
\cite{7,11,14}.

This theory has drawn significant attention and some cosmological
features have been studied comprehensively. We have investigated the
validity of first and second laws of thermodynamics in $f(R,T)$
gravity. It is shown that equilibrium picture of thermodynamics may
not be achieved due to matter geometry interaction \cite{18}. The
reconstruction of $f(R,T)$ Lagrangian is executed under various
considerations likewise, considering an auxiliary scalar field
\cite{19}, family of holographic DE models in the background of FRW
universe \cite{20} and anisotropic solutions \cite{21}. Alvarenga
\cite{22} discussed scalar matter perturbations for a particular
model which assures the standard continuity equation and obtained
matter density perturbed equations.

It is shown that for $f(R,T)$ Lagrangian unlikely consequences are obtained
from quasistatic approximation as compared to those derived in agrement with
the $\Lambda$CDM model. Jamil et al. \cite{23} explored the reconstruction of
$f(R,T)$ theory corresponding to cosmological solutions like $\Lambda$CDM,
phantom as well as non-phantom matter fluids and Einstein static universe.
However, they used the standard continuity equation without any additional
constraint on $f(R,T)$ gravity \cite{22}. We have used a significant approach
for cosmological reconstruction in terms of e-folding reproducing different
cosmological eras and the stability of $f(R,T)$ models is also analyzed
\cite{24}.

Recently, a more complicated modified theory is developed \cite{25,26} which
involves the nonminimal coupling through contraction of the Ricci and
energy-momentum tensors referred to $f(R,T,R_{\mu\nu}T^{\mu\nu})$ gravity
whose action is the extended Lagrangian of $f(R,T)$ gravity. It would be
interesting to explore different cosmic features in this theory. The
classical energy conditions of general relativity (GR) are profound to the
Hawking-Penrose singularity theorems and classical black hole laws of
thermodynamics \cite{27}. These conditions have been used to address several
important issues in GR and cosmology \cite{28}. Energy conditions have been
investigated in modified theories with different considerations such as
$f(R)$ gravity \cite{29}, $f(R)$ gravity with nonminimal coupling to matter
\cite{30}, $f(R,\mathcal{L}_m)$ gravity \cite{31}, $f(T)$ gravity \cite{32},
scalar-tensor theory \cite{32a}, modified Gauss-Bonnet gravity \cite{33} and
$f(R,T)$ gravity \cite{34}. We have also discussed the energy conditions in
$f(R,T)$ gravity and investigated the stability of power law solutions for
particular class of models \cite{34}.

In this work, we are interested to develop the energy conditions
bounds in $f(R,T,R_{\mu\nu}T^{\mu\nu})$ gravity and analyze some
specific models. The energy conditions permit to set bounds for
attractiveness property of gravity as well as energy density being
positive. The paper has the following format. In the next section,
the fundamental formulation of the field equations is presented.
Section \textbf{3} comprises a brief review of energy conditions in
GR and also the respective inequalities in this modified theory. In
section \textbf{4}, we consider some specific forms of
$f(R,T,R_{\mu\nu}T^{\mu\nu})$ gravity and illustrate the energy
conditions bounds as well as analyze the Dolgov-Kawasaki
instability. Section \textbf{5} concludes our findings.

\section{$f(R,T,R_{\mu\nu}T^{\mu\nu})$ Gravity}

The $f(R,T,R_{\mu\nu}T^{\mu\nu})$ gravity is an interesting candidate among
the modified theories which are based on nonminimal coupling between matter
and geometry. The action of this modified theory is of the form \cite{25,26}
\begin{equation}\label{1}
\mathcal{A}=\frac{1}{2{\kappa}^2}\int{dx^4\sqrt{-g}\left[f(R,T,R_{\mu\nu}T^{\mu\nu})
+\mathcal{L}_{m}\right]},
\end{equation}
where $\kappa^2=1$, $f(R,T,R_{\mu\nu}T^{\mu\nu})$ is an arbitrary function in
all of its contents, the Ricci scalar $R$, trace of the energy-momentum
tensor $T=T^\mu_\mu$ and contraction of the Ricci tensor with $T_{\mu\nu}$,
$\mathcal{L}_{m}$ denotes the Lagrangian density of matter part. The matter
energy-momentum tensor is given by \cite{35}
\begin{equation}\label{2}
T_{{\mu}{\nu}}=-\frac{2}{\sqrt{-g}}\frac{\delta(\sqrt{-g}
{\mathcal{\mathcal{L}}_{m}})}{\delta{g^{{\mu}{\nu}}}}.
\end{equation}
If the matter action depends only on the metric tensor rather than on its
derivatives then the energy-momentum tensor yields
\begin{equation}\label{3}
T_{{\mu}{\nu}}=g_{{\mu}{\nu}}\mathcal{L}_{m}-\frac{2{\partial}
{\mathcal{L}_{m}}}{\partial{g^{{\mu}{\nu}}}}.
\end{equation}
The field equations in $f(R,T,R_{\mu\nu}T^{\mu\nu})$ gravity can be found by
varying the action (\ref{1}) with respect to $g_{\mu\nu}$ as
\begin{eqnarray}\nonumber
&&R_{{\mu}{\nu}}f_{R}-\{\frac{1}{2}f-\mathcal{L}_{m}f_T-\frac{1}{2}\nabla_\alpha
\nabla_\beta(f_QT^{\alpha\beta})\}g_{{\mu}{\nu}}+(g_{{\mu}{\nu}}
{\Box}-{\nabla}_{\mu}{\nabla}_{\nu})f_{R}\\\nonumber&+&\frac{1}{2}
\Box(f_QT_{{\mu}{\nu}})+2f_QR_{\alpha(\mu}T^\alpha_{\nu)}-\nabla_\alpha\nabla
_{(\mu}[T^\alpha_{\nu)}f_Q]-G_{\mu\nu}\mathcal{L}_{m}f_Q-2\left(f_Tg^{\alpha\beta}
\right.\\\label{4}&+&\left.f_QR^{\alpha\beta}\right)\frac{\partial^2\mathcal
{L}_{m}}{\partial{g}^{\mu\nu}\partial{g}^{\alpha\beta}}=(1+f_T+\frac{1}
{2}R{f}_Q)T_{\mu\nu},
\end{eqnarray}
where we set $Q=R_{\mu\nu}T^{\mu\nu}$ to make the equations more convenient
while the subscripts indicate the derivatives with respect to $R,~T$ and $Q$.

One can obtain the field equations in $f(R,T)$ and $f(R)$ theories
from the above expression by substituting some particular functions
of Lagrangian. For vacuum, this leads to the field equations in
$f(R)$ gravity. The field equation (\ref{4}) can be rearranged in
the following form
\begin{equation}\label{5}
G_{\mu\nu}=R_{{\mu}{\nu}}-\frac{1}{2}Rg_{{\mu}{\nu}}=T_{{\mu}{\nu}}^{eff},
\end{equation}
which is analogous to the standard field equations in GR. Here
$T_{{\mu}{\nu}}^{eff}$, the effective energy-momentum tensor in
$f(R,T,Q)$ gravity is defined as
\begin{eqnarray}\nonumber
{T}_{{\mu}{\nu}}^{eff}&=&\frac{1}{f_{R}-f_Q\mathcal{L}_{m}}\left[(1+f_T+
\frac{1}{2}R{f}Q)T_{\mu\nu}+\{\frac{1}{2}(f-R{f}_R)-\mathcal{L}_{m}f_T
\right.\\\nonumber&-&\left.\frac{1}{2}\nabla_\alpha
\nabla_\beta(f_QT^{\alpha\beta})\}g_{{\mu}{\nu}}-(g_{{\mu}{\nu}}
{\Box}-{\nabla}_{\mu}{\nabla}_{\nu})f_{R}-\frac{1}{2}
\Box(f_QT_{{\mu}{\nu}})\right.\\\label{6}
&-&\left.2f_QR_{\alpha(\mu}T^\alpha_{\nu)}+
\nabla_\alpha\nabla_{(\mu}[T^\alpha_{\nu)}f_Q]+2\left(f_Tg^{\alpha\beta}
+f_QR^{\alpha\beta}\right)\frac{\partial^2\mathcal{L}_{m}}{\partial{g}^{\mu\nu}
\partial{g}^{\alpha\beta}}\right].
\end{eqnarray}
Applying the covariant divergence to the field equation (\ref{4}),
we obtain
\begin{eqnarray}\nonumber
\nabla^{\mu}T_{\mu\nu}&=&\frac{2}{2(1+f_T)+R{f}_Q}
\left[\nabla_\mu(f_QR^{\alpha\mu}T_{\alpha\nu})+\nabla_\nu(\mathcal{L}_mf_T)
-\frac{1}{2}(f_Q{R}_{\sigma\zeta}\right.\\\label{7}&+&\left.f_T{g}_{\sigma\zeta})
\nabla_\nu{T}^{\sigma\zeta}-G_{\mu\nu}\nabla^\mu(f_Q\mathcal{L}_m)-\frac{1}{2}
\left[\nabla^\mu(R{f}_Q)+2\nabla^\mu{f}_T\right]T_{\mu\nu}\right].
\end{eqnarray}
It is significant to see that ideal continuity equation does not agree in
this modified theory which is also true in other modified theories involving
non-minimal matter geometry coupling \cite{7}-\cite{15}.

\section{Energy Conditions}

\subsection{Raychaudhuri Equation}

To discuss the energy conditions in modified theories one needs to adopt the
procedure originally developed in Einstein gravity. We first discuss these
conditions in GR and search a way to express them in this modified theory. In
fact, Raychaudhuri equation plays a key role to prove singularity theorems
and explain the congruence of timelike and null geodesics. Raychaudhuri's
equation for the congruence of timelike geodesics is defined as \cite{36}
\begin{equation}\label{8}
\frac{d{\theta}}{d\tau}=-\frac{1}{3}{\theta}^2-{\sigma}^{\mu\nu}{\sigma}_{\mu\nu}
+{\omega}^{\mu\nu}{\omega}_{\mu\nu}-R_{\mu\nu}u^{\mu}u^{\nu},
\end{equation}
where $\theta$ denotes the expansion parameter (if $\theta>0$ then congruence
will be diverging and for $\theta<0$, it will be converging),
${\sigma}_{\mu\nu}$ and ${\omega}_{\mu\nu}$ measure the distortion of volume
and rotation of curves linked to the congruence set by the vector field
$u^\mu$. In case of null geodesics characterized by the vector field
$\kappa^\mu$, the temporal variation of expansion is given by
\begin{equation}\label{9}
\frac{d{\theta}}{d\tau}=-\frac{1}{2}{\theta}^2-{\sigma}^{\mu\nu}{\sigma}_{\mu\nu}
+{\omega}^{\mu\nu}{\omega}_{\mu\nu}-R_{\mu\nu}\kappa^{\mu}\kappa^{\nu}.
\end{equation}
It is significant to remark that Raychaudhuri equation is exclusively
geometric and hence develops no deal to any theory of gravity under
discussion. Actually, the energy-momentum tensor can have contribution from
different sources and it is convenient to set some constraints to deal it on
physical grounds. There are certain inequalities which may limit the
arbitrariness in the energy-momentum tensor based on Raychaudhuri equation
with attractiveness property of gravity. The association of Raychaudhuri
equation can be set from the fact that the variation of expansion parameter
is related to $T_{\mu\nu}$ if one finds the Ricci tensor from the field
equations. Hence, one can develop the physical constraints on the
energy-momentum tensor through the connection between Raychaudhuri equation
and the field equations.

As ${\sigma}^{\mu\nu}{\sigma}_{\mu\nu}\geqslant0$ (shear tensor is purely
spatial), so for any hypersurface orthogonal congruence
($\omega_{\mu\nu}=0$), the condition of attractive gravity takes the form
\begin{equation}\label{10}
\textbf{SEC}: \quad R_{\mu\nu}u^{\mu}u^{\nu}\geqslant0, \quad
\textbf{NEC}: \quad R_{\mu\nu}\kappa^{\mu}\kappa^{\nu}\geqslant0.
\end{equation}
Using the field equations, one can relate $R_{\mu\nu}$ to $T_{\mu\nu}$ so
that the above conditions become
\begin{equation}\label{11}
R_{\mu\nu}u^{\mu}u^{\nu}=(T_{\mu\nu}-\frac{T}{2}g_{\mu\nu})
u^{\mu}u^{\nu}\geqslant0, \quad
R_{\mu\nu}\kappa^{\mu}\kappa^{\nu}=T_{\mu\nu}\kappa^{\mu}
\kappa^{\nu}\geqslant0.
\end{equation}
If the matter part is considered as perfect fluid
\begin{equation}
T_{{\mu}{\nu}}=({\rho}+p)u_{\mu}u_{\nu}-pg_{{\mu}{\nu}},
\end{equation}
where energy density and pressure are denoted by $\rho$ and $p$,
then these conditions reduce to the most familiar form of strong and
null energy conditions in GR,
\begin{equation}\label{12}
\rho+3p\geqslant0, \quad \quad \rho+p\geqslant0.
\end{equation}

\subsection{Energy Conditions in $f(R,T,Q)$ Gravity}

Here, we adopt the procedure developed in \cite{30,31} for
$f(R),~f(R,\mathcal{L}_m)$ and $f(R)$ gravity with arbitrary and nonminimal
matter geometry coupling to extend it to a more general $f(R,T,Q)$ grvaity.
The Ricci tensor in Eq.(\ref{5}) can be represented in terms of
$T_{\mu\nu}^{eff}$ and its trace $T^{eff}$ as
\begin{equation}\label{13}
R_{{\mu}{\nu}}=T_{{\mu}{\nu}}^{eff}-\frac{1}{2}g_{{\mu}{\nu}}T^{eff},
\end{equation}
where the contraction of Eq.(6) yields the trace of the energy-momentum
tensor
\begin{eqnarray}\nonumber
T^{eff}&=&\frac{1}{f_{R}-f_Q\mathcal{L}_{m}}\left[(1+f_T+
\frac{1}{2}R{f}Q)T+2(f-R{f}_R)-4\mathcal{L}_{m}f_T\right.\\\nonumber&-&\left.
\nabla_\alpha\nabla_\beta(f_QT^{\alpha\beta})-3{\Box}f_{R}-\frac{1}{2}\Box(f_QT)
-2f_{Q}R_{\alpha\beta}T^{\alpha\beta}+2g^{\mu\nu}(f_Tg^{\alpha\beta}\right.
\\\label{14}&+&\left.f_QR^{\alpha\beta})\frac{\partial^2\mathcal{L}_{m}}
{\partial{g}^{\mu\nu}\partial{g}^{\alpha\beta}}\right].
\end{eqnarray}
The attractive nature of gravity needs to satisfy the following additional
constraint
\begin{equation}\label{15}
\frac{1+f_T+\frac{1}{2}R{f}_Q}{f_R-f_Q\mathcal{L}_m}>0
\end{equation}
which does not depend on the conditions (\ref{10}) derived from the
Raychaudhuri equation. In fact this condition corresponds to the effective
gravitational coupling in $f(R,T,Q)$ gravity.

We take the homogeneous and isotropic flat FRW metric defined as
\begin{equation}\nonumber
ds^{2}=dt^2-a^2(t)d\textbf{x}^2,
\end{equation}
where $a(t)$ represents the scale factor and $d\textbf{x}^2$ is the spatial
part of the metric. The corresponding effective energy density and pressure
can be taken such that $T_{\mu\nu}^{eff}$ assumes the form of perfect fluid.
In FRW background, $\rho_{eff}$ and $p_{eff}$ can be obtained in this
modified theory as
\begin{eqnarray}\nonumber
\rho_{eff}&=&\frac{1}{f_{R}-f_Q\mathcal{L}_{m}}\left[\rho+(\rho
-\mathcal{L}_m)f_T+\frac{1}{2}(f-R{f}_R)-3H\partial_t{f}_R-\frac{3}{2}(3H^2\right.
\\\label{16}
&-&\left.\dot{H})\rho{f}_Q-\frac{3}{2}(3H^2+\dot{H})pf_Q+\frac{3}{2}H\partial_t
[(p-\rho)f_Q]\right],
\end{eqnarray}
\begin{eqnarray}\nonumber
{p}_{eff}&=&\frac{1}{f_{R}-f_Q\mathcal{L}_{m}}\left[p+(p
+\mathcal{L}_m)f_T+\frac{1}{2}(R{f}_R-f)+\frac{1}{2}(\dot{H}+3H^2)\rho{f}_Q
\right.\\\nonumber&+&\left.\frac{1}{2}(3H^2-\dot{H})p{f}_Q+\partial_{tt}{f}_R
+2H\partial_tf_R+\frac{1}{2}\partial_{tt}[(\rho-p)f_Q]+2H\partial_t
[(\rho\right.\\\label{17}&+&\left.p){f}_Q]\right],
\end{eqnarray}
where $R=-6(\dot{H}+2H^2),~H=\frac{\dot{a}}{a}$ being Hubble
parameter and over dot refers to time derivative. Here, we neglect
the terms involving second derivative of matter Lagrangian with
respect to the metric tensor. As we are dealing with perfect fluid,
so matter Lagrangian can either be $\mathcal{L}_m=p$ or
$\mathcal{L}_m=-\rho$ which makes it obvious to ignore such term.

In this modified theory, we can employ an approach analogous to that in GR
and combine Eqs.(\ref{11}) and (\ref{13}) so that SEC is of the form
\begin{equation}\label{18}
T^{eff}_{\mu\nu}u^\mu{u}^\nu-\frac{1}{2}T^{eff}\geqslant0,
\end{equation}
where $g_{\mu\nu}u^\mu{u}^\nu=1$. Using Eqs.(\ref{6}) and (\ref{14}), it
follows that
\begin{eqnarray}\nonumber
\rho_{eff}+3p_{eff}&=&\frac{1}{f_{R}-f_Q\mathcal{L}_{m}}\left[(\rho+3p)+(\rho+3p
+2\mathcal{L}_m)f_T+R{f}_R-f\right.\\\nonumber&+&\left.3[\dot{H}(\rho-p)]f_Q
+3H\partial_t[f_R+\frac{1}{2}(3\rho+5p)f_Q]\right.\\\label{19}
&+&\left.3\partial_{tt}[f_R+\frac{1}{2}(\rho-p)f_Q]\right]\geqslant0,
\end{eqnarray}
which is the SEC in $f(R,T,Q)$ gravity. One can represent the NEC in
$f(R,T,Q)$ gravity in the form
\begin{equation}\nonumber
T_{\mu\nu}^{eff}\kappa^{\mu}\kappa^{\nu}\geqslant0.
\end{equation}
Inserting Eq.(\ref{6}) in the above relation results the following inequality
\begin{eqnarray}\nonumber
\rho_{eff}+p_{eff}&=&\frac{1}{f_{R}-f_Q\mathcal{L}_{m}}\left[(1+f_T)(\rho+p)
-3H^2(\rho+p)f_Q+2\dot{H}(\rho-p)f_Q\right.\\\label{20}&-&\left.H\partial_t\{f_R-\frac{1}{2}
(\rho+7p)f_Q\}+\partial_{tt}\{f_R+\frac{1}{2}(\rho-p)f_Q\}\right]\geqslant0.
\end{eqnarray}
It is remarked that one can obtain the NEC and SEC in $f(R)$ and $f(R,T)$
modified theories by taking $f(R,T,Q)=f(R)$ and $f(R,T,Q)=f(R,T)$,
respectively. Moreover, the traditional structures for the NEC
($\rho+p\geqslant0$) and SEC ($\rho+3p\geqslant0$) can be found in the
framework of GR as a specific case with $f(R,T,Q)=R$. In determining the WEC
and DEC, we consider the modified form of energy conditions in GR which are
obtained under the transformations $\rho\rightarrow\rho_{eff}$ and
$p\rightarrow{p}_{eff}$. We would like to mention here that the null and
strong energy conditions given by Eqs.(\ref{19}) and (\ref{20}) are derived
from the Raychaudhuri equation. One can obtain equivalent results following
the same procedure as that in GR with conditions
$\rho_{eff}+p_{eff}\geqslant0$ and $\rho_{eff}+3p_{eff}\geqslant0$.

We extend this approach to develop the constraints for WEC and DEC
so that these conditions for $f(R,T,Q)$ gravity are given by
$\rho_{eff} \geqslant0$ and $\rho_{eff}-p_{eff}\geqslant0$. Using
Eqs.(\ref{16}) and (\ref{17}), we can obtain the constraints on WEC
and DEC. The WEC requires the condition (\ref{20}) and the following
inequality
\begin{eqnarray}\nonumber
\rho_{eff}&=&\frac{1}{f_{R}-f_Q\mathcal{L}_{m}}\left[\rho+(\rho
-\mathcal{L}_m)f_T+\frac{1}{2}(f-R{f}_R)-3H\partial_t{f}_R-\frac{3}{2}(3H^2\right.
\\\label{21}
&-&\left.\dot{H})\rho{f}_Q-\frac{3}{2}(3H^2+\dot{H})pf_Q+\frac{3}{2}H\partial_t
[(p-\rho)f_Q]\right]\geqslant0,
\end{eqnarray}
whereas the DEC is satisfied by meeting the inequalities (\ref{20}),
(\ref{21}) and the condition
\begin{eqnarray}\nonumber
\rho_{eff}-p_{eff}&=&\frac{1}{f_{R}-f_Q\mathcal{L}_{m}}\left[(\rho-p)+(\rho-p
-2\mathcal{L}_m)f_T+f-R{f}_R\right.\\\nonumber&+&\left.\{\dot{H}(\rho-p)-6H^2(\rho+p)\}f_Q
-H\partial_t[\frac{1}{2}(7\rho+p)f_Q+5f_R]\right.\\\label{22}&-&\left.\partial_{tt}[f_R+
\frac{1}{2}(\rho-p)f_Q]\right]\geqslant0.
\end{eqnarray}
When we take $f(R,T,Q)=f(R,T)$, the above expressions reduce to the
WEC and DEC in $f(R,T)$ gravity which are similar to that in
\cite{34}. Also, by neglecting the dependence on the trace of
energy-momentum tensor, we can have the energy conditions in $f(R)$
gravity which are consistent with the results in \cite{29}. If the
variation of Lagrangian with respect to $T$ and $Q$ is null then
such conditions constitute $\rho\geqslant0$ and $\rho+p\geqslant0$,
i.e., the WEC and DEC in GR.

One can utilize the energy conditions constraints (\ref{19})-(\ref{22}) to
restrict some specific models in $f(R,T,Q)$ gravity in the framework of FRW
metric. To be more definite about these energy constraints, we define
deceleration, jerk and snap parameters as \cite{37}
\begin{eqnarray}\nonumber
q=-\frac{1}{H^2}\frac{\ddot{a}}{a}, \quad
j=\frac{1}{H^3}\frac{\dddot{a}}{a}, \quad \text{and} \quad
s=\frac{1}{H^4}\frac{\ddddot{a}}{a}.
\end{eqnarray}
and express the Hubble parameter and its time derivatives in terms of these
parameters
\begin{eqnarray}\nonumber
&&\dot{H}=-H^2(1+q), \quad \ddot{H}=H^3(j+3q+2), \\\nonumber
&&\dddot{H}=-H^4(5q+2j-s+3).
\end{eqnarray}
Since $R$, $\dot{R}$ and $\ddot{R}$ are represented in terms of the above
relations, so using these parameters the energy conditions
(\ref{19})-(\ref{22}) can be constituted as
\begin{eqnarray}\nonumber
&&(\rho+p)(1+f_T)+\frac{1}{2}\{\ddot{\rho}-\ddot{p}+H(\dot{\rho}+7\dot{p})
-4(1+q)H^2(\rho-p)-6H^2\\\nonumber&\times&(\rho+p)\}f_Q-6H^2(s-j+(q+1)(q+8))f_{RR}+(\ddot{T}
-H\dot{T})f_{RT}+\{\ddot{Q}\\\nonumber&-&H\dot{Q}-3H^3(j-q-2)(2(\dot{\rho}-\dot{p})
+H(\rho+7p))-3H^4(\rho-p)(s+q^2+8q\\\nonumber&+&6)\}f_{RQ}+\frac{1}{2}\{2\ddot{T}
+(2(\dot{\rho}-\dot{p})+H(\rho+7p))\dot{T}\}f_{TQ}+\frac{1}{2}\{2\ddot{Q}+
(2(\dot{\rho}-\dot{p})\\\nonumber&+&H(\rho+7p))\dot{Q}\}f_{QQ}+[6H^3(j-q-2)]^2
f_{RRR}-12H^3(j-q-2)\dot{T}f_{RRT}\\\nonumber&+&\{18(\rho-p)[H^3(j-q-2)]^2-12H^3
(j-q-2)\dot{Q}\}f_{RRQ}+2\dot{T}[\dot{Q}-3H^3(\rho\\\nonumber&-&p)(j-q-2)]f_{RTQ}
+\dot{T}^2f_{RTT}+\dot{Q}[\dot{Q}-6H^3(\rho-p)(j-q-2)]f_{RQQ}+
\frac{1}{2}\\\nonumber&\times&(\rho-p)\dot{T}[\dot{Q}f_{TQQ}+\dot{T}f_{TTQ}]
+\frac{1}{2}(\rho-p)\dot{Q}[\dot{T}f_{TQQ}+\dot{Q}f_{QQQ}]\geqslant0,
~\textbf{(NEC)}\\\label{23}\\\nonumber
&&\rho(1+f_T)-\mathcal{L}_mf_T+\frac{1}{2}f+3H^2(1-q)f_{R}+\frac{3}{2}H\{\dot{p}
-\dot{\rho}-2H(2\rho+p)\\\nonumber&-&H(\rho-p)q\}f_Q-3H\{\dot{Q}+3H^3(p-\rho)
(j-q-2)\}f_{RQ}+\frac{3}{2}H(p-\rho)(\dot{Q}\\\label{24}&\times&f_{QQ}+\dot{T}
f_{TQ})+18H^4(j-q-2)f_{RR}-3H\dot{T}f_{RT})\geqslant0,~~~\textbf{(WEC)}
\end{eqnarray}
\begin{eqnarray}\nonumber
&&(\rho+3p)(1+f_T)+2\mathcal{L}_mf_T-f-6H^2(1-q)f_{R}+\frac{3}{2}\{\ddot{\rho}
-\ddot{p}+H(3\dot{\rho}+5\dot{p})\\\nonumber&+&6H^2[(p-\rho)(1+q)]\}f_Q-18H^4
(s+j+q^2+7q+4)f_{RR}+3(\ddot{T}\\\nonumber&+&H\dot{T})f_{RT}+3\{\ddot{Q}+H\dot{Q}
-3H^3(j-q-2)[2(\dot{\rho}-p)+H(3\rho+5p)]-3H^4\\\nonumber&\times&(\rho-p)(s+q^2
+8q+6)\}f_{RQ}+\frac{3}{2}\{(\rho-p)\ddot{T}+[2(\dot{\rho}-\dot{p})+H(3\rho+5p)
\dot{T}]\}\\\nonumber&\times&f_{TQ}+\frac{3}{2}\{(\rho-p)\ddot{Q}+[2(\dot{\rho}
-\dot{p})-H(\rho+3p)\dot{Q}]\}f_{QQ}+3[6H^3(j-q-2)]^2\\\nonumber&\times&f_{RRR}
-36H^3(j-q-2)\dot{T}f_{RRT}+3\{-12H^3(j-q-2)\dot{Q}+18(\rho-p)[6\\\nonumber
&\times&H^3(j-q-2)]^2\}f_{RRQ}+6\dot{T}\{\dot{Q}-3(\rho-p)H^3(j-q-2)\}f_{RTQ}
+3\dot{T}^2\\\nonumber&\times&f_{RTT}+3\dot{Q}\{\dot{Q}-6 H^3(\rho-p)(j-q-2)\}
f_{RQQ}+\frac{3}{2}(\rho-p)\dot{T}\{2\dot{Q}f_{TQQ}\\\label{25}&+&\dot{T}f_{TTQ}
\}+\frac{3}{2}(\rho-p)\dot{Q}^2f_{QQQ}\geqslant0,~~~\textbf{(SEC)},
\end{eqnarray}
\begin{eqnarray}\nonumber
&&(\rho-p)(1+f_T)-2\mathcal{L}_mf_T+f+6H^2(1-q)f_{R}+\frac{1}{2}\{\ddot{p}
-\ddot{\rho}-H(7\dot{p}+\dot{\rho})\\\nonumber&-&12H^2(\rho+p)-2H^2(\rho-p)(1+q)\}
f_Q-\{\ddot{Q}+5H\dot{Q}-6H^3(j-q-2)\\\nonumber&\times&(\dot{\rho}-\dot{p})-3H^4
(j-q-2)(7\rho+p)-3H^4(s+q^2+8q+6)(\rho-p)\}\\\nonumber&-&\frac{1}{2}\{(\rho-p)\ddot{T}
+[2(\dot{\rho}-\dot{p})+H(7\rho+p)]\dot{T}\}f_{TQ}-\frac{1}{2}
\{(\rho-p)\ddot{Q}+[2(\dot{\rho}-\dot{p})\\\nonumber&+&H(7\rho+p)]
\dot{Q}\}f_{QQ}+6H^4[s+5j+(q-1)(q+4)]f_{RR}-(\ddot{T}+5\dot{T}H)f_{RT}\\\nonumber&-&[6H^3
(j-q-2)]^2f_{RRR}+12H^3(j-q-2)\dot{T}f_{RRT}+\{12H^3(j-q-2)\dot{Q}
\\\nonumber&-&18[H^3(j-q-2)]^2(\rho-p)\}f_{RRQ}-2\dot{T}\{\dot{Q}-6H^3(j-q-2)
(\rho-p)\}f_{RTQ}\\\nonumber&-&\dot{T}^2f_{RTT}-\dot{Q}\{\dot{Q}-6H^3(j-q-2)
(\rho-p)\}f_{RQQ}-\frac{1}{2}(\rho-p)\dot{T}\{\dot{Q}f_{TQQ}\\\label{26}&+&\dot{T}f_{TTQ}\}
\frac{1}{2}(\rho-p)\dot{Q}\{\dot{T}f_{TQQ}+\dot{Q}f_{QQQ}\}
\geqslant0.~~~\textbf{(DEC)}
\end{eqnarray}
The results of energy conditions in terms of cosmographic parameters for
$f(R)$ and $f(R,T)$ theories can be achieved from the constraints
(\ref{22})-(\ref{26}).

\section{Constraints on Class of $f(R,T,Q)$ Models}

To illustrate how these energy conditions put limits on $f(R,T,Q)$ gravity,
we consider some specific functional forms for the Lagrangian (\ref{1})
namely \cite{26},
\begin{enumerate}
\item $f(R,T,Q)=R+\alpha{Q}$,
\item $f(R,T,Q)=R(1+\alpha{Q})$,
\end{enumerate}
where $\alpha$ is a coupling parameter. Recently, these models have been
studied in \cite{26} which suggest that exponential and de Sitter type
solutions exist for these forms of $f(R,T,Q)$ gravity. Thus one can deduce
that coupling between matter and geometry may cause the current cosmic
acceleration.

\subsection{$f(R,T,Q)=R+\alpha{Q}$}

In the first place, we consider the Lagrangian given by $R+\alpha{Q}$. In FRW
background, the energy conditions for such model can be represented as
\begin{equation}\label{27}
\alpha{A_1}+H\partial_tA_2\geqslant{A_3},
\end{equation}
where $A_i$'s purely depend on the energy conditions under discussion. For
NEC, one can have
\begin{eqnarray}\nonumber
A_1^{NEC}&=&(2\dot{H}-3H^2)\rho-(2\dot{H}+3H^2)p+\partial_{tt}[\alpha^{-1}+\frac{1}{2}(\rho-p)],\\\label{28}
A_2^{NEC}&=&-(1-\frac{\alpha}{2}(\rho+7p)), \quad A_3^{NEC}=-(\rho+p).
\end{eqnarray}
For WEC, this yields
\begin{eqnarray}\label{29}
A_1^{WEC}&=&-3H^2\rho,\quad A_2^{WEC}=\frac{3\alpha}{2}(p-\rho)-3,\quad
A_3^{WEC}=-\rho.
\end{eqnarray}
For SEC, one can find
\begin{eqnarray}\nonumber
A_1^{SEC}&=&3\rho(2\dot{H}+H^2)-3p(2\dot{H}+3H^2)+3\partial_{tt}[\alpha^{-1}+\frac{1}{2}
(\rho-p)],\\\label{30}
A_2^{SEC}&=&3(1+\frac{\alpha}{2}(3\rho+5p)), \quad A_3^{SEC}=-(\rho+3p).
\end{eqnarray}
For DEC, it follows that
\begin{eqnarray}\nonumber
A_1^{DEC}&=&2\dot{H}(p-\rho)+6H^2(p-\rho)-\partial_{tt}[\alpha^{-1}+\frac{1}{2}
(\rho-p)],\\\label{31}
A_2^{DEC}&=&-\frac{\alpha}{2}(7\rho+p)-5, \quad A_3^{DEC}=-(\rho-p).
\end{eqnarray}

We can also find the condition of attractive gravity for this model
from inequality (\ref{27}) so that
\begin{eqnarray}\nonumber
A_1^{AG}=\left(1-\alpha{\mathcal{L}_m}\right)\left(\frac{1}{\alpha}+\frac{R}{2}
\right)^{-1},\quad A_2^{AG}=constant, \quad A_3^{AG}=0.
\end{eqnarray}
The energy conditions (\ref{27})-(\ref{31}) can be expressed in
terms of deceleration parameter (see appendix A.1).  It can be seen
that these conditions depend only on the parameters $H,~q$ and
$\alpha$. In our discussion, we set the present day values of
cosmographic parameters as
$q_0=-0.81^{+0.14}_{-0.14},~j_0=2.16^{+0.81}_{-0.75}$ \cite{38} and
$H_0=73.8$ \cite{39}, while matter is assumed to be pressureless. To
exemplify how these conditions can constrain the above model, we
consider the WEC given by the relation
\begin{equation}\label{31a}
\rho-6H^2-3\alpha{H}\dot{\rho}\geqslant0.
\end{equation}
For the given $H$ and $q$, one can see that the above inequality relies on
the measures of parameter $\alpha$ and time derivative of energy density.
Here, $\dot{\rho}$ can be evaluated using Eq.(\ref{7}) which takes the form
\begin{equation}\nonumber
\dot{\rho}=-6H\rho\{1+\alpha{H}^2(2+5q-3H(1+q))\}/\{2-3\alpha(2-q)H^2\}.
\end{equation}
It can be seen that $\dot{\rho}$ is always negative. Using this value of
$\dot{\rho}$ in Eq.(\ref{31a}), we find that WEC for the model
$f(R,T,Q)=R+\alpha{R}$ is satisfied if $\alpha>0$ for present day values of
$q$ and $H$.

\subsection{$f(R,T,Q)=R(1+\alpha{Q})$}

In second example, we consider the function $f$ given by
$R(1+\alpha{Q})$ and energy conditions for such model can be written
as
\begin{equation}\label{32}
\hat{\alpha}{B_1}+H\partial_tB_2\geqslant{B_3},
\end{equation}
where $\hat{\alpha}=(-1)\alpha$ and $B_i$'s purely depend on the energy
conditions under discussion. For NEC, one can have
\begin{eqnarray}\nonumber
B_1^{NEC}&=&[-3H^2(\rho+p)+2\dot{H}(\rho-p)]R+\partial_{tt}[\hat{\alpha}^{-1}
+Q+\frac{1}{2}(\rho-p)R]
,\\\label{33} B_2^{NEC}&=&-(1+\hat{\alpha}{Q}-\frac{\hat{\alpha}}{2}(\rho+7p)R),
\quad B_3^{NEC}=-(\rho+p).
\end{eqnarray}
For WEC, one can have
\begin{eqnarray}\nonumber
B_1^{WEC}&=&-\frac{3}{2}[(3H^2-\dot{H})\rho+(3H^2+\dot{H})p]R,\\\label{34}
B_2^{WEC}&=&3[\frac{\hat{\alpha}}{2}(p-\rho)R-(1+\hat{\alpha}{Q})],
\quad B_3^{WEC}=-\rho.
\end{eqnarray}
For SEC, it follows that
\begin{eqnarray}\nonumber
B_1^{SEC}&=&3[\dot{H}(\rho-p)]R+\partial_{tt}[\hat{\alpha}^{-1}+Q+\frac{1}{2}
(\rho-p)R],\\\label{35}
B_2^{SEC}&=&3[1+\hat{\alpha}{Q}+\frac{\hat{\alpha}}{2}(3\rho+5p)R],
\quad B_3^{SEC}=-(\rho+3p).
\end{eqnarray}
For DEC, this yields
\begin{eqnarray}\nonumber
B_1^{DEC}&=&-6H^2(\rho+p)R-(\rho+p)\dot{H}R-\partial_{tt}[\hat{\alpha}^{-1}
+Q+\frac{1}{2}(\rho-p)R],\\\label{36}
B_2^{DEC}&=&-(5(1+\alpha{Q})+\frac{\alpha}{2}(7\rho+p)R), \quad B_3^{DEC}=-(\rho-p).
\end{eqnarray}
The condition of attractive gravity can be obtained from the inequality
(\ref{32}) and relevant components are
\begin{eqnarray}\nonumber
B_1^{AG}=\left(1+\hat{\alpha}{Q-R\mathcal{L}_m}\right)\left(\frac{1}{\hat{\alpha}}
+\frac{R^2}{2}\right)^{-1}, \quad B_2^{AG}=constant, \quad B_3^{AG}=0.
\end{eqnarray}

The viability of modified theories is under debate to develop the
criteria for different modifications to the Einstein-Hilbert action.
In this perspective, one of the important criterion is
Dolgov-Kawasaki instability which has been developed to constrain
the $f(R)$ gravity and $f(R)$ gravity with curvature matter coupling
\cite{40}. Recently, the authors \cite{25,26} have executed this
instability analysis for $f(R,T,Q)$ gravity which yields the
condition of Dolgov-Kawasaki instability as
\begin{equation}\label{37}
3f_{RR}+\left(\frac{1}{2}T-T^{00}\right)f_{QR}\geqslant0.
\end{equation}
For the model $f=R(1+\alpha{Q})$, the inequality (\ref{37}) takes the form
\begin{equation}\nonumber
\alpha(\rho-3p)+6\hat{\alpha}{H}(\rho+p)\partial_t\left(\frac{H}{R}\right)\geqslant0,
\end{equation}
where
\begin{equation}\nonumber
\hat{\alpha}= \begin{cases} (-1)\alpha & \text{if } {R,Q}<0 \\
\alpha & \text{if } {R,Q}>0 \end{cases}.
\end{equation}
One can derive the above inequality using the relation (\ref{32}) so that
$B_i$'s are given by
\begin{eqnarray}\nonumber
B_1^{AG}=\frac{\rho-3p}{\rho+p}, \quad B_2^{AG}=\frac{6\hat{\alpha}H}{R}
, \quad B_3^{AG}=0.
\end{eqnarray}
We check the validity of constraints (\ref{32})-(\ref{36}) for this model.
The constraint to ensure WEC is given by
\begin{equation}\nonumber
\rho[1+9\alpha{H}^4(2j-q^2-3q+2)]+9\alpha{H}^3(1-2q)\dot{\rho}\geqslant0.
\end{equation}
As in the previous case, we evaluate $\dot{\rho}$
\begin{equation}\nonumber
\dot{\rho}=-3H\rho\{1+6\alpha{H^4}(j-4q+2q^2-1)\}/\{1+9\alpha{H}^4(1-q)(2-q)\}.
\end{equation}
Here, $\dot{\rho}<0$ for any value of $\alpha$ and hence the WEC is satisfied
only if parameter $\alpha$ is positive.

\subsection{$f(R,T)$ Models}

Here we present $f(R,T,Q)$ gravity models which involve null
variation with respect to $Q$ and corresponds to $f(R,T)$ gravity.
Although the energy conditions are examined in $f(R,T)$ gravity but
the constraints are developed for very particular cases. Alvarenga
et al. \cite{41} studied the energy conditions for some models of
the type $f(R,T)=R+2f(T)$ and analyse their stability under matter
perturbations. In our previous work \cite{34}, we established the
energy condition constraints for those $f(R,T)$ models which confirm
the existence of power law solutions in this modified theory. Here
we are interested in more general functional forms of $f(R,T)$
involving an exponential function and also the coupling between $R$
and $T$. We present the energy condition constraints for the
following models
\begin{enumerate}
\item $f(R,T)=\alpha{\exp}\left(\frac{R}{\alpha}+\lambda{T}\right)$
\item $f(R,T)=R+\eta{R}^mT^n$
\end{enumerate}
where $\alpha$, $\lambda$, $\eta$, $m$ and $n$ are arbitrary constants.
\begin{itemize}
\item $f(R,T)=\alpha{\exp}\left(\frac{R}{\alpha}+\lambda{T}\right)$
\end{itemize}
If $\frac{R}{\alpha}+\lambda{T}\ll1$ then
$f(R,T)\approx\alpha+R+\frac{\lambda}{\alpha}T+...$ representing the
$\Lambda$CDM model. The energy constraints in $f(R,T)$ gravity can
be achieved by placing null variation of $f$ with respect to $Q$ in
the results (\ref{19})-(\ref{22}) which are similar to that in \cite{34}.

For the exponential model, these conditions take the form
\begin{equation}\label{38}
\frac{\exp\left(\frac{R}{\alpha}+\lambda{T}\right)}{1+\alpha\lambda\exp
\left(\frac{R}{\alpha}+\lambda{T}\right)}\left(C_1+C_2\right)\geqslant{C_3},
\end{equation}
where $C_i$'s depend on the energy conditions given in Appendix \ref{A2}. The
condition of attractive gravity in $f(R,T)$ gravity is $(1+f_T)/f_R>0$ which
becomes $(1+\alpha\lambda\exp\left(\frac{R}{\alpha}+\lambda{T}\right))/\exp
\left(\frac{R}{\alpha}+\lambda{T}\right)>0$ for the exponential model. We can
obtain this inequality from Eq.(\ref{38}) for $C_1=1$, $C_2=C_3=0$. It is
suggested \cite{26} that Dolgov-Kawasaki instability in $f(R,T)$ gravity
would be identical to that in $f(R)$ gravity so that one can check the
viability of $f(R,T)$ models on similar steps as in $f(R)$ theory. Thus for
$f(R,T)$ theory, we have
\begin{equation}\nonumber
f_R(R,T)>0,\quad f_{RR}(R,T)>0,\quad R{\geq}R_{0}.
\end{equation}
For this model, the instability conditions are formulated as
$\exp\left(\frac{R}{\alpha}+\lambda{T}\right)>0$ and
$\frac{1}{\alpha}\exp\left(\frac{R}{\alpha}+\lambda{T}\right)>0$ which can be
derived from relation (\ref{38}) by taking
$C_1=1,~C_2=\alpha\lambda\exp\left(\frac{R}{\alpha}+\lambda{T}\right),~C_3=0$
and $C_1+C_2=\frac{1}{\alpha}\left(1+\alpha\lambda\exp\left(\frac{R}{\alpha}
+\lambda{T}\right)\right)$,\\~$C_3=0$, respectively. One can represent energy
conditions (\ref{A2}) in the form of cosmographic parameters. It is mentioned
here that the measurement of present day value of snap parameter `$s$' has
not been reported from reliable sources so far.

The energy conditions corresponding to the above model depend on `$s$' except
the WEC, so we explore the validity of WEC. The inequality to fulfill the WEC
is given by
\begin{eqnarray}\nonumber
&&\rho\left(\frac{1+\alpha\lambda\exp\left(\frac{R}{\alpha}+\lambda{T}\right)}
{\exp\left(\frac{R}{\alpha}+\lambda{T}\right)}\right)+\alpha(0.5-\lambda
{\mathcal{L}_m})+3H^2\{(1-q)+6\alpha^{-1}H^2\\\nonumber&\times&(j-q-2)\}
-3\lambda{H}\dot{T}\geqslant0.
\end{eqnarray}
Using the WEC results in GR, i.e., $\rho>0$ and also the condition
of attractive gravity
$(1+\alpha\lambda\exp\left(\frac{R}{\alpha}+\lambda{T}\right))/\exp
\left(\frac{R}{\alpha}+\lambda{T}\right)>0$, the above inequality is
reduced to
\begin{eqnarray}\nonumber
&&\alpha(0.5-\lambda{\mathcal{L}_m})+3H^2\{(1-q)+6\alpha^{-1}H^2(j-q-2)\}
-3\lambda{H}\dot{T}\geqslant0.
\end{eqnarray}
We set the matter Lagrangian density as $\mathcal{L}_m=p$ and assume the
pressureless matter so that
\begin{eqnarray}\label{39}
&&0.5\alpha+3H^2\{(1-q)+6H^2(j-q-2)\alpha^{-1}\}-3\lambda{H}\dot{\rho}
\geqslant0.
\end{eqnarray}
If we consider the present day values of the parameters like Hubble,
deceleration and jerk then the above inequality depends on $\dot\rho$ and
values of constants $(\alpha,\lambda)$. We find $\dot{\rho}$ from the energy
conservation equation in $f(R,T)$ gravity
\begin{equation}\label{40}
\dot{\rho}+3H(\rho+p)=\frac{-1}{1+f_{2T}}\left[(\rho-\mathcal{L}_m)
\dot{f_{T}}-\dot{\mathcal{L}_m}f_{T}+\frac{1}{2}\dot{T}f_{T}\right].
\end{equation}

For exponential model, this takes the form
\begin{equation}\nonumber
\dot{\rho}=-\frac{3H\rho\{1+\lambda(\alpha-2H^2(j-q-2))\exp
\left(\frac{R}{\alpha}+\lambda{T}\right)\}}{1+(1.5+\lambda\rho)\alpha\lambda
\exp\left(\frac{R}{\alpha}+\lambda{T}\right)}
\end{equation}
If $\alpha>0$ then the first two terms in inequality (\ref{39}) are
positive whereas for the last term we need to have $-\dot\rho>0$.
From the above expression, we see that $-\dot\rho>0$ if $\lambda>0$
and $\alpha>2H_0^2(j_0-q_0-2)$. Thus, the WEC for exponential
$f(R,T)$ model is satisfied if $\lambda>0$ and
$\alpha>2H_0^2(j_0-q_0-2)$. We consider another form of matter
Lagrangian $\mathcal{L}_m=-\rho$ for which the continuity equation
and constraint to fulfill the WEC are given by
\begin{eqnarray}\nonumber
\dot{\rho}=-\frac{3H\rho\{1+\lambda(\alpha-4H^2(j-q-2))\exp
\left(\frac{R}{\alpha}+\lambda{\rho}\right)\}}{1+(1.5+2\lambda\rho)\alpha\lambda
\exp\left(\frac{R}{\alpha}+\lambda{\rho}\right)},\\\nonumber
\alpha(0.5+\lambda\rho)+3H^2\{(1-q)+6\alpha^{-1}H^2(j-q-2)\}
-3\lambda{H}\dot{\rho}\geqslant0.
\end{eqnarray}
As in the previous case, we find a constraint for which $\dot\rho<0$ which is
only possible if $\alpha>4H^2(j-q-2)$. It is to be noted that we set the
present values of $H$ and other parameters so that the WEC is satisfied if
$\lambda>0$ and $\alpha>4H^2(j-q-2)$.
\begin{itemize}
\item $f(R,T)=R+\eta{R}^mT^n$
\end{itemize}

Here, we consider the power law type $f(R,T)$ model which involves coupling
between $R$ and $T$. Such functional form of $f(R,T)$ matches to the form of
Lagrangian $f(R,T)=f_1(R)+f_2(R)f_3(T)$ with $f_1(R)=R,f_2(R)=R^m$ and
$f_3(T)=T^n$ which involves the explicit nonminimal gravitational matter
geometry coupling. In a recent work \cite{24}, we have reconstructed such
type of $f(R,T)$ models corresponding to power law solutions. The
attractiveness of gravity implies that
$\{1+\hat\eta{n}R^mT^{n-1}\}/\{1+\hat\eta{m}R^{m-1}T^n\}>0$. The energy
condition constraints for this model can be represented as
\begin{equation}\label{41}
\frac{\hat\eta|R|^m|T|^n}{1-\hat\eta{n}|R|^m|T|^{n-1}}[D_1+
\mathcal{L}_mT^{-1}D_2]\geqslant{D}_3,
\end{equation}
where $D_i$'s can have particular relations depending on the energy
conditions which are shown in appendix \ref{A3}.

We are concerned to study the WEC inequality for this model and
develop the constraints as in exponential model. The condition to
meet the WEC is given by
\begin{eqnarray}\nonumber
&&\rho+\eta{R}^m\rho^n\left[n(1-\mathcal{L}_m\rho^{-1})+\frac{1}{2}(1-m)
-3m(m-1)H\dot{R}R^{-2}\right.\\\nonumber&-&\left.3mnH\dot{\rho}R^{-1}
\rho^{-1}\right]\geqslant0.
\end{eqnarray}
Initially, we consider $\mathcal{L}_m=p$ for which the above inequality can
be represented in the form of deceleration and jerk parameters as
\begin{eqnarray}\nonumber
&&2\rho+2\eta[6H^2(1-q)]^m\rho^n\left[2n+1-m+m(m-1)\frac{j-q-2}
{(1-q)^2}\right.\\\label{42}&+&\left.\frac{mn\dot{\rho}\rho^{-1}}
{2(1-q)H}\right]\geqslant0.
\end{eqnarray}
The coupling parameter $\eta$ is assumed to be positive so that the above
constraint is satisfied if one can meet the condition in square bracket. For
this purpose, $\dot{\rho}$ can be obtained using Eq.(\ref{40}) in the form
\begin{equation}\nonumber
\dot\rho=-\frac{3H\rho\{1+n\eta[6H^2(1-q)]^m\rho^{n-1}\left(1+\frac{m(j-q-2)}
{3(1-q)}\right)\}}{1+n(n+0.5)\eta[6H^2(1-q)]^m\rho^{n-1}}.
\end{equation}
Substituting $\dot{\rho}$ in Eq.(\ref{42}), it is found that WEC is
satisfied if both the constants $m$ and $n$ are positive. One can
also examine the WEC constraint for $\mathcal{L}_m=-\rho$ for which
WEC can be met if $m,n>0$ with coupling parameter being positive.

\section{Conclusions}

The late time accelerated cosmic expansion is a major issue in cosmology.
Modified gravity has appeared as a handy candidate to address such issues
which predict the destiny of the universe. Recently, a more generalized
modified theory is established on the basis of curvature matter coupling
named as $f(R,T,R_{\mu\nu}T^{\mu\nu})$ gravity \cite{25,26}. This theory is
an extensive form of $f(R,T)$ gravity where Lagrangian not only depends on
$R$ and $T$  but also involves the contribution from contraction of the Ricci
and the energy-momentum tensors. In this modified theory, extra force is
always present due to the coupling between matter and geometry leading to
motion of test particles as nongeodesic. In previous studies of nonminimal
coupling \cite{14,15}, it has been encountered that extra force disappears if
one chooses $\mathcal{L}_m=-p$ whereas it does not vanish even for this
$\mathcal{L}_m$ in $f(R,T,R_{\mu\nu}T^{\mu\nu})$ theory. The expression of
extra force in $f(R,T,R_{\mu\nu}T^{\mu\nu})$ gravity \cite{26} explicitly
depends on $R_{\mu\nu}$ as compared to $f(R,T)$ gravity. Thus the deviation
from geodesic motion can be more significant in this case.

Lagrangian of $f(R,T,R_{\mu\nu}T^{\mu\nu})$ gravity is more comprehensive
implying that different functional forms of $f$ can be suggested. The
versatility in Lagrangian raises the question how to constrain such theory on
physical grounds. In \cite{25,26}, the authors have studied the validity of
this theory and developed the notable Dolgov-Kawasaki instability (the
condition of stability against local perturbations). In this paper, we have
developed some constraints on general as well as specific forms of
$f(R,T,R_{\mu\nu}T^{\mu\nu})$ gravity by examining the respective energy
conditions. The NEC and SEC are derived using the Raychaudhuri equation along
with the condition that gravity is attractive. These conditions are more
general as compared to those derived in $f(R)$ and $f(R,T)$ theories.
Moreover, these inequalities are equivalent to the results found from
conditions $\rho+3p\geqslant0$ and $\rho+p\geqslant0$ under the
transformations $\rho\rightarrow\rho_{eff}$ and $p\rightarrow{p}_{eff}$,
respectively. One can employ the similar procedure to derive the WEC and DEC
by translating their counterpart in GR for effective energy-momentum tensor.
The conditions of positive effective gravitational coupling and attractive
nature of gravity are also obtained in this theory.

To illustrate how these conditions can constrain the
$f(R,T,R_{\mu\nu}T^{\mu\nu})$ gravity, we have taken two functional forms of
$f$ namely, $f=R+\alpha{Q}$ and $f=R(1+\alpha{Q})$. It is shown that WEC for
these models depends on the coupling parameter $\alpha$ and is satisfied only
if $\alpha$ is positive. We have also set the Dolgov-Kawasaki criterion in
this discussion. In section \textbf{4.3}, the $f(R,T)$ gravity is addressed
as a specific case to this modified theory. We have taken two interesting
choices for the Lagrangian, one involving an exponential function and other
having explicit coupling between $R$ and $T$. The validity of WEC for both
choices of matter Lagrangian $\mathcal{L}_m=p$ and $\mathcal{L}_m=-\rho$ have
been explored. The WEC for
$f(R,T)=\alpha{\exp}\left(\frac{R}{\alpha}+\lambda{T}\right)$ is met in both
cases if coupling parameter $\lambda>0$ and $\alpha>2H_0^2(j_0-q_0-2)$(or
$\alpha>4H_0^2(j_0-q_0-2)$), respectively. For the model
$f(R,T)=R+\eta{R}^mT^n$, the WEC is satisfied if both the constants $m$ and
$n$ are positive.

\vspace{.25cm}

{\bf Acknowledgment}

\vspace{.25cm}

The authors thank the Higher Education Commission, Islamabad, Pakistan for
its financial support through the \emph{Indigenous Ph.D. 5000 Fellowship
Program Batch-VII}.

\renewcommand{\theequation}{A.\arabic{equation}}
\setcounter{equation}{0}
\section*{Appendix A}

\begin{eqnarray}\nonumber
\textbf{NEC}&:&\rho+p+\frac{\alpha}{2}\{\ddot{\rho}-\ddot{p}+H(\dot{\rho}+7\dot{p})
-H^2[\rho(5+q)+p(1-2q)]\}\geqslant0,\\\nonumber
\textbf{WEC}&:&\rho+\frac{3\alpha}{2}\{H(\dot{p}-\dot{\rho})-2H^2\}\geqslant0,\\\nonumber
\textbf{SEC}&:&\rho+3p+\frac{3\alpha}{2}\{\ddot{\rho}-\ddot{p}+H(3\dot{\rho}
+5\dot{p})-2H^2(1+2q)\rho-2(1-2q)\\\nonumber&\times&H^2p\}\geqslant0,\\\nonumber
\textbf{DEC}&:&\rho-p+\frac{\alpha}{2}\{\ddot{p}-\ddot{\rho}-H(7\dot{\rho}
+\dot{p})+4H^2\rho(q-2)\rho+4H^2(2-q)\\\label{A1}&\times&p\}\geqslant0.
\end{eqnarray}
\begin{eqnarray}\nonumber
\textbf{NEC}&:&C_1^{NEC}=\frac{1}{\alpha^2}\left(\dot{R}^2+\alpha
(\ddot{R}-H\dot{R})+2\alpha\lambda\dot{R}\dot{T}\right)+\lambda\{\ddot{T}-H\dot{T}
+\lambda\dot{T}^2\},
\\\nonumber &&C_2^{NEC}=0, \quad C_3^{NEC}=-(\rho+p),\\\nonumber
\textbf{WEC}&:&C_1^{WEC}=3(\dot{H}+2H^2)-\frac{3}{\alpha}H\dot{R}
-3\lambda{H}\dot{T},~C_2^{WEC}=\alpha\left(\frac{1}{2}-\lambda{\mathcal{L}}
_m\right),\\\nonumber && C_3^{WEC}=-\rho,
\\\nonumber
\textbf{SEC}&:&C_1^{SEC}=R+\frac{3}{\alpha^2}\left(\dot{R}^2+\alpha
(\ddot{R}+H\dot{R})+2\alpha\lambda\dot{R}\dot{T}\right)+3\lambda\{\ddot{T}+\dot{T}
(H\\\nonumber&+&\lambda\dot{T})\},\quad
C_2^{SEC}=\alpha(2\lambda{\mathcal{L}}_m-1), \quad C_3^{NEC}=-(\rho+3p),
\\\nonumber
\textbf{DEC}&:&C_1^{DEC}=-R-\frac{1}{\alpha^2}\left(\dot{R}^2+\alpha
(\ddot{R}+5H\dot{R})+2\alpha\lambda\dot{R}\dot{T}\right)-\lambda\{\ddot{T}+\dot{T}
(5H\\\label{A2}&+&\lambda\dot{T})\},\quad
C_2^{DEC}=\alpha(1-2\lambda{\mathcal{L}}_m), \quad C_3^{DEC}=-(\rho-p).
\end{eqnarray}
\begin{eqnarray}\nonumber
\textbf{NEC}&:&D_1^{NEC}=m(m-1)R^{-2}\{\ddot{R}-H\dot{R}+(m-2)\dot{R}^2R^{-1}
+2n\dot{R}\dot{T}T^{-1}\}\\\nonumber&+&nR^{-1}T^{-1}\{\ddot{T}-H\dot{T}+(n-1)
\dot{T}^2T^{-1}\},\quad D_2^{NEC}=0,
\\\nonumber &&D_3^{NEC}=-(\rho+p),\\\nonumber
\textbf{WEC}&:&D_1^{WEC}=(1-m)\{0.5+3mH\dot{R}R^{-2}\}-3mnH\dot{T}R^{-1}T^{-1}
,\\\nonumber && D_2^{WEC}=-n,\quad D_3^{WEC}=-\rho
\\\nonumber
\textbf{SEC}&:&D_1^{SEC}=(m-1)\{1+3mR^{-2}\{\ddot{R}+H\dot{R}+(m-2)\dot{R}^2R^{-1}
+2n\dot{R}\dot{T}\\\nonumber&\times&T^{-1}\}\}+3mnR^{-1}T^{-1}\{\ddot{T}+H\dot{T}+(n-1)
\dot{T}^2T^{-1}\},\quad D_2^{SEC}=2n,
\\\nonumber&&D_3^{SEC}=-(\rho+3p),
\\\nonumber
\textbf{DEC}&:&D_1^{DEC}=(1-m)\{1+mR^{-2}\{\ddot{R}+5H\dot{R}+(m-2)\dot{R}^2R^{-1}
+2n\dot{R}\dot{T}\\\nonumber&\times&T^{-1}\}\}-mnR^{-1}T^{-1}\{\ddot{T}+5
H\dot{T}+(n-1)T^{-1}\},\quad D_2^{DEC}=-2n,
\\\label{A3} && D_3^{DEC}=-(\rho-p).
\end{eqnarray}

\end{document}